# Upper Bounds of Interference Alignment Degree of Freedom

Feng Ouyang, Senior Member, IEEE

*Abstract*— Interference alignment allows multiple users to share the same frequency and time resource in a wireless communications system. At present, two performance bounds, in terms of degree of freedom, have been proposed. One is for infinite-dimension extension and the other is for MIMO systems. This paper provides an understanding of the MIMO bound by examining its proofs and shows that it does not apply to a more practical case: MIMO-OFDM. Several approaches are proposed in searching for DoF bounds for systems such as finite-dimension time extension and MIMO-OFDM systems.

## I. INTRODUCTION

Interference alignment (IA) is a technique that allows multiple users (transmit/receive pairs) to use the same frequency and time resource in a wireless communications system, while controlling the mutual interference in order to achieve maximum aggregated channel capacity [1]. In this context, the channel capacity is measured under the limit of high signal to noise ratio (SNR) and expressed in terms of degree of freedom (DoF). IA techniques are especially valuable to modern cellular communications systems, where, due to smaller cell sizes, mutual interference, rather than thermal noise, is the limiting factor for system capacity.

IA is a precoding technique. Simply put, the transmitters, based on the present channel conditions, carefully choose the transmit signal design to achieve the following effect. At each receiver, all interferences overlap with each other, leaving some room for the interference-free reception of desired signal. The detailed formulation will be presented in Section II.

### A. Two performance bounds

An IA scheme was proposed in 2008 by Cadambe and Jafar [2]. The work focused on single-antenna users, but can be easily extended to multiple-antenna (multiple-input-multiple-output, or MIMO) users. In both single-antenna and multiple-antenna cases, the scheme uses time-extension. Namely, a symbol is transmitted over a number of time slots, without cross-interference among the time slots. The channel gains among all pairs of transmitters and receivers (including both desired and interference ones) change independently among all channels and from one time slot to the next. An explicit solutions for precoder design was provided in [2], demonstrating the potential benefit in overall capacity. Based on such solution, a tight upper performance bound was put forward (referred to as Bound A hereafter) as the signal dimension (in this case the number of time slots per symbol) approaches *infinity*. However, in order to realize a significant performance gain, the time-extension must be massive. As will be shown in Section III, the number of time slots for each symbol increases exponentially with the square of the number of users and can easily be as large as millions. This results in significant implementation problems in terms of complexity and latency.

Yeits, et al. studied another scheme, which applies IA to MIMO systems without time-extension [3]. In this case, all symbols are transmitted within one time slot. IA is achieved in the vector space spanned by the multiple antennas (the same vector space used in conventional MIMO transmission). Without providing explicit solutions, this work put forward another upper bound (referred to as Bound B hereafter) in achievable performance. There are several works subsequently reaffirming such bound and finding sufficient conditions for the IA solutions [4, 5, and 6]. Such analysis approach was extended to a combination of MIMO and time/frequency-extension [7, 8].

All above methods assume that all channel state information (CSI) is available at a central controller, which designs the precoding solutions for all transmitters. More practical methods have been explored, where the precoding designs are optimized in distributed fashions [9, 10, 11, 12, and 13]. In this paper, we focus on the centralized optimization scheme studied by [2] and [3].

### B. The Question

In recent years, most of the works on IA are based on Bound B. It states that the capacity of an IA system is approximately twice the capacity of a simple time division multiple access (TDMA) or frequency division multiple access (FDMA) medium sharing system. In particular, per-user capacity is inversely proportional to the number of users. While still significant, such improvement is incremental and may be discouraging to the IA research community. Furthermore, it is qualitatively lower than Bound A, which states that as the number of users increase, per-user capacity is maintained at a constant level. Bound A implies that the

This material is based in part upon work supported by the National Science Foundation of the United States, under Grant No. CCF-0964495.

The author is with Johns Hopkins University, Applied Physics Laboratory, Laurel, MD 20723 USA (e-mail: feng.ouyang@jhuapl.edu).

aggregated capacity increases as the number of users increase. The performance difference between the two bounds is very significant as the number of users increases in an IA scheme.

The original scheme proposed in [2] uses time-extension. This requires the knowledge of future channel states in order to design the precoding coefficients. A perhaps more practical way is using OFDM modulation, where the number of subcarriers constitutes the signal dimensions (assuming each sub-channel fades independently) [7]. This can be further combined with MIMO [8]. However, the maximum practical numbers for OFDM subcarriers are in thousands for most of today's communications system (such as LTE). Therefore, one cannot practically reach Bound A using OFDM scheme, based on the IA solution provided in [2]. The question is: can an IA system still do better than Bound B in the case of limited extensions such as an OFDM system with single or multiple antennas?

This paper identifies some gaps in our understanding of the performance bounds and proposes possible approaches to address them. It is believed that such efforts help establish performance bounds for practical IA systems based on OFDM.

The rest of the paper is organized as follows. Section II presents the problem statement and mathematical formulation. Section III outlines the performance upper bounds presented in current literature. It shows the contradiction between Bound A and Bound B. One of the proofs of Bound B is recounted in more details in Section IV, where it is pointed out that such proof does not apply to the case of diagonal channel matrices (as in the time-extension or OFDM cases). Section V discusses several possible approaches to further study the problem. The paper is then summarized in Section VI.

## II. SYSTEM CONFIGURATION AND PROBLEM STATEMENT

### A. System Configuration and IA Formulation

Consider $K$ pairs of users sharing the same wireless channel. For IA considerations, noise is ignored. Therefore, the received signal is

$$y^{[j]} = \sum_{k=1}^{K} H^{[j,k]} x^{[k]}. \tag{1}$$

$x^{[k]}$ is the transmit signal. It is a vector of size $N_k$ in a linear space:

$$x^{[k]} = \begin{pmatrix} x_1^{[k]} \\ x_2^{[k]} \\ \vdots \\ x_{N_k}^{[k]} \end{pmatrix} \tag{2}$$

The components of $x^{[k]}$ can be signals applied to different antennas (in MIMO case), sent at different time slots (in time-extension case), or sent at different subcarriers (in single-antenna OFDM case), etc. $N_k$ is the signal space dimension for user $k$. Its meaning will be illustrated later with some examples. $y^{[j]}$ is a vector of size $N_j$, representing received signal for user $j$. $H^{[j,k]}$ is a matrix representing the vector channel from the $k$th transmitter to the $j$th receiver. It is of size $N_j \times N_k$. More discussions about $H^{[j,k]}$ will be provided later in this section. For user $j$, $x^{[j]}$ is the desired signal and all other terms in the summation are interferences.

In an IA scheme, the transmit signal is precoded as:

$$x^{[k]} = V^{[k]} s^{[k]}. \tag{3}$$

Here $s^{[k]}$ is a vector symbol (of size $d_k$), to be transmitted by user $k$. $d_k$ is the number of the scalar symbols that can be transmitted simultaneously by user $k$. $V^{[k]}$ is the precoding matrix of size $N_k \times d_k$. It converts data vector $s^{[k]}$ to signal vector $x^{[k]}$.

On the receiver side, the recovered data is

$$\hat{s}^{[k]} = U^{[k]H} y^{[k]}. \tag{4}$$

$U^{[k]}$ is the decoding matrix of size $N_k \times d_k$. $(\cdot)^H$ represents Hermitian operation to the matrix. The decoding matrix $U^{[k]}$ converts received signal $y^{[k]}$ to data vector $\hat{s}^{[k]}$ (of size $d_k$).

In order to achieve IA, we wish that $\hat{s}^{[k]}$ contains full information about the desired signal $s^{[k]}$. Namely, we wish that $U^{[k]} H^{[k,k]} V^{[k]}$ is a full rank matrix (but not necessarily an identity matrix). At the same time, all interferences should not be present in $\hat{s}^{[k]}$. Therefore, IA presents the following requirements:

$$\left(U^{[j]}\right)^H H^{[j,k]} V^{[k]} = 0, \forall j \neq k, \tag{5}$$

$$\text{Rank}\left[\left(U^{[k]}\right)^H H^{[k,k]} V^{[k]}\right] = d_k, \forall k. \tag{6}$$

Depending on the particular setups, $H^{[k,j]}$ may subject to more constraints. In the case of single antenna, time-extension schemes studied in [2], components of signal are spread over $N_s$ time slots. Each time slot is assumed to have a different channel gain (as in fast fading situation). There is no cross interference among signals transmitted in different time slots. Therefore, in this case $H^{[j,k]}$ are all diagonal matrices with a common size $N_k = N_s$. If signal extension is performed across subcarrier frequencies in an OFDM system [7], $H^{[j,k]}$ are diagonal, as well. Here $N_s$ would be the number of subcarriers used for signal extension.

In the case of MIMO without other extensions [3], $H^{[j,k]}$ is the channel matrix across all transmit antennas and receive antennas, just as in a typical MIMO problem setup. $N_k = M_k$, where $M_k$ is the number of antennas for user $k$. (It is assumed that for each user pair, the numbers of transmit antennas and receive antennas are the same.) In this case $H^{[j,k]}$ are non-spares, i.e., none of its elements is confined to zero.

In the case of MIMO with frequency-extension (MIMO-OFDM) [8], the signals are spread over both antennas and subcarriers. There is cross interference among signals transmitted by different antennas at the same subcarrier. However, there is no cross interference among different subcarriers. Therefore, $H^{[j,k]}$ is a block-diagonal matrix. Each block has the size of $M_j \times M_k$. There are $N_c$ such blocks, representing $N_c$ subcarriers. $N_k = M_k N_c$.

In IA problems, the channel matrix $H^{[j,k]}$ is assumed to be generic. Namely, all elements (except those confined to zero) are drawn independently from a statistical process (usually uniform distribution). This implies that the channel is uncorrelated and full rank in MIMO case, and has independent fading among frequency and time slots in OFDM and time-extension cases.

Table 1 summarizes the parameters used in this paper to describe the system. Index $k$ denotes the users. Table 2 summarizes the vectors and matrices used in this paper.

TABLE 1
SUMMARY OF PARAMETERS

| Symbol | Meaning |
|---|---|
| $K$ | Number of users |
| $N_k$ | Size of the signal vector for user $k$ |
| $d_K$ | Size of the data vector for user $k$ |
| $M_k$ | Number of antennas for user $k$ |
| $N_c$ | Number of subcarriers |
| $N_s$ | Number of time slots |

TABLE 2
VECTORS AND MATRICES IN IA FORMULATION

| Symbol | Size | Meaning |
|---|---|---|
| $s^{[k]}$ | $d_k$ | Tx data |
| $x^{[k]}$ | $N_k$ | Tx signal |
| $V^{[k]}$ | $N_k \times d_k$ | Precoder |
| $H^{[k,j]}$ | $N_k \times N_j$ | Channel |
| $U^{[k]}$ | $N_k \times d_k$ | Decoder |
| $y^{[k]}$ | $N_k$ | Rx signal |
| $\hat{s}^{[k]}$ | $N_k$ | Rx data |

The problem of IA is stated as follows. Given all parameters in Table 1 and channel matrices $\{H^{[j,k]}\}$, find precoding and combining matrices $\{U^{[k]}\}$ and $\{V^{[k]}\}$ to satisfy equations (5) and (6).

Note that if $\{U^{[k]}, V^{[k]}\}$ is a solution to (5), then $\{U^{[k]}A^{[k]}, V^{[k]}B^{[k]}\}$ is also a solution, for any full rank matrices $A^{[k]}$ and $B^{[k]}$ of size $d_k \times d_k$. This provides a freedom to set $d_k^2$ elements in each of the $U^{[k]}$ and $V^{[k]}$. For example, the first $d_k$ rows of $U^{[k]}$ and $V^{[k]}$ can be set to be identity matrices. Such choice guarantees equation (6) to be satisfied. Therefore, from now on equation (5) is the focus of analysis.

*B. Degree of Freedom*

Performance of an IA system is characterized by the degree of freedom (DoF). It is defined as [2]:

$$d \stackrel{\text{def}}{=} lim_{\rho \to \infty} \frac{R(\rho)}{log(\rho)}. \tag{7}$$

Here $\rho$ is the SNR, which does not include mutual interference. $R$ is the data rate. Comparing to Shannon's law, $d$ is the equivalent number of independent data streams one can transmit independently. In the system defined in Section II.A, $d$ is the sum of $d_k$ over all users, where $d_k$ can be considered as DoF for user $k$.

This paper focuses on the upper bound of $d$ that yield non-zero solutions $\{U^{[k]}, V^{[k]}\}$ for equations (5) and (6), given all other parameters.

## III. CURRENT DoF UPPER BOUND RESULTS

In this section, relevant results in the literature are recaptured to provide a framework for further discussion.

*A. Single Antenna Case (Bound A)*

For single antenna systems with time-extension, all $H^{[j,k]}$ matrices are diagonal and all $N_k$ are equal. Namely, $N_k = N_s \ \forall k$. In this case, [2] provided explicit solutions to equation (5) for some specific parameter sets. For these solutions, an integer $N$ is defined as:

$$N \stackrel{\text{def}}{=} (K-1)(K-2) - 1. \tag{8}$$

Here $K$ is the number of users. The signal space dimension is

$$N_s = (n+1)^N + n^N. \tag{9}$$

Here $n$ is an arbitrary positive integer. Under such signal space dimension, the DoFs for the users are

$$d_1 = (n+1)^N, d_k = n^N \ \forall k \neq 1. \tag{10}$$

The total DoF is

$$d \stackrel{\text{def}}{=} \sum_{k=1}^{K} d_k = n^N(K-1) + (n+1)^N. \tag{11}$$

The normalized DoF (i.e., DoF per time slot, per user) is

$$\bar{d} \stackrel{\text{def}}{=} \frac{d}{KN_s} = \frac{(K-1)n^N + (n+1)^N}{K[(n+1)^N + n^N]}. \tag{12}$$

It is easy to see that

$$\bar{d} < \frac{1}{2}, \quad lim_{n \to \infty} \bar{d} = \frac{1}{2}. \tag{13}$$

The asymptotic performance is very attractive: each user can send 0.5 data symbols per time slot, regardless of the number of users. In contrast, in a conventional medium sharing scheme such as TDMA, the data rate for each user is inversely proportional to the number of users $K$. Without any overhead, $\bar{d} = \frac{1}{K}$ in TDMA case.

It can be argued that $\bar{d} = \frac{1}{2}$ is an absolute upper bound [2]. The explicit solutions provided by [2] show such bound is reachable, at least asymptotically. On the other hand, these solutions apply to only a small set of signal space dimension $N_s$ values given by equation (9). For other $N_s$ values, solutions to equation (5) may exist, but they are not given in this scheme. Furthermore, when $n$ is not infinity, the solutions provided by [2] are not necessarily optimal (i.e., giving the largest possible $\bar{d}$). Therefore, with finite signal dimensions, the upper bound of $\bar{d}$ was not given by this scheme.

*B. MIMO Case (Bound B)*

For MIMO systems, IA is performed across the space dimension. $H^{[j,k]}$ is none-sparse (i.e., none of its elements is confined to zero), and $N_k$ is the number of antennas for user $k$. This case was studied by [3] and others. A necessary condition for the existence of nonzero solutions to equation (5) was proposed by [3]: the number of free variables must be no less than the number of equations. This is referred to as the "properness" requirement.

For (5), the number of equations is

$$N_e = \sum_{k \neq l} d_k d_l . \tag{14}$$

To get the number of variables, it was noted in [3] (and explained in Section II.A) that any valid solution can be transformed to another valid solution, where the first $d_i$ rows of $U^{[i]}$ and $V^{[i]}$ form identity matrices. Therefore, there are actually $N_i d_i - d_i^2$ free variables in $U^{[i]}$ and $V^{[i]}$ each. Such choice also ensures that equation (6) is automatically satisfied. The number of variables is thus

$$N_v = 2 \sum_k (N_k d_k - d_k^2) . \tag{15}$$

Therefore, the "properness" condition is

$$N_e \leq N_v . \tag{16}$$

Or,

$$\sum_{k \neq l} d_k d_l - 2 \sum_k (N_k d_k - d_k^2) \leq 0 . \tag{17}$$

On the other hand, properness is not a sufficient condition for solutions. References [3-6] further studied sufficient conditions (referred to as "feasibility") for the existence of solutions. Feasibility is out of the scope of this paper.

Consider the case where all users have the same number of antennas and the same DoF. Namely,

$$N_k = M, d_k = \bar{d} . \tag{18}$$

Inequality (17) becomes

$$\bar{d} \leq \frac{2M}{K+1} . \tag{19}$$

If K users, each with $M$ antennas working in the MIMO spatial multiplexing mode, share the medium in a TDMA fashion, then each user will have a DoF of $M / K$. Therefore, IA approximately doubles the capacity. The DoF of each user is still inversely proportional to the number of users. The benefit of IA, as predicted by (17), is much smaller than those predicted by (13) for the single antenna, time-extension case.

*C. MIMO OFDM case (Bound B)*

For a MIMO OFDM system, IA can be extended across the subcarriers. This has been examined using the same "properness" requirement [7, 8]. It was found that such extension does not result in significant additional benefit comparing to the MIMO case in the last section, if the "properness" requirement is upheld. For the case of MIMO OFDM, there is an opportunity of further reducing the number of free variables [7, 8]. However, such reduction is minor and does not change the overall picture. Therefore, it is not discussed in detail here.

*D. The Apparent Contradiction between the Two Bounds*

As described in the preceding subsections, Bound B is supported by a general argument. Therefore, one would expect that it covers the case of single antenna, time-extension IA schemes. However, although [2] does not provide DoF upper bound with finite signal dimensions, it provides explicit solutions that, with some choice of parameters, clearly yields DoF that is higher than Bound B. Therefore, there appears to be a contradiction.

Therefore, an understanding of this issue may be helpful in devising an IA scheme based on OFDM or MIMO-OFDM that yields performance exceeding Bound B. This issue will be further discussed in Section IV, where proof of Bound B will be described in more details.

IV. PROOF OF THE "PROPERNESS" CONDITION AND THE DIAGONAL MATRICES

This section looks more closely into the proofs of the properness condition that leads to the DoF bound B, inequality (17), and discuss why such bound does not apply to the cases of diagonal matrices and block-diagonal matrices.

The initial argument of the properness condition is based on the premise that the number of equations cannot be more than the number of variables, for an equation system to have meaningful solutions [3]. Such requirement is valid for linear equations that are linear independent. It is also is intuitively reasonable. However, such requirement does not hold in general for polynomial equations. Therefore, we need to prove that the properness requirement indeed holds for the IA equations (5). There are two such proofs given in the literature [3, 6]. They will be examined in more details in the following subsections. It will be shown that these proofs do not apply to the cases of time-extension and MIMO-OFDM, where channel matrices are spares (i.e., some elements are confined to zero).

*A. Proof in [3], Based on Bernstein's Theorem*

In [3], Bernstein's theorem was invoked to support the properness as a necessary condition for the existence of none-trivial solutions to equation (5). Without going into the details, the arguments can be outlined as follows.

Bernstein's theorem, roughly speaking, says that if there are $N_e$ polynomial equations with $N_v$ variables, and the coefficients of these equations are generic, then there are at most a finite number of common solutions when $N_e = N_v$.[1]

If $N_e > N_v$ (i.e., when properness is not held), then the finite number of solutions to the $N_v$ equations must also satisfy the additional $N_e - N_v$ equations. Since these equations have generic coefficients, this is impossible. Therefore, properness ($N_e \leq N_v$) is a necessary condition for the existence of solutions.

However, as noted by Sun and Luo [14], Bernstein's theory and the above corollary apply only to "non-zero" solutions, i.e., in the context of equation (5), none of the elements in $\{U^{[i]}, V^{[j]}\}$ can be zero. Obviously, such constraint does not exist in the original IA problem.

This problem can be solved in the case of MIMO IA, which was the subject of [3], as explained in [14]. It can be shown that for MIMO IA, if a system has a general solution, it must have a "zone-zero" solution. Therefore, the necessary condition for the general solution is the same as that for the non-zero solution. However, such correspondence requires the equations (i.e., the $\{H^{[i,j]}\}$ matrices) to be generic and structureless. For example, if all $\{H^{[i,j]}\}$ are diagonal, non-existence of non-zero solutions does not imply non-existence of general solutions. This explains the difference between Bound A and Bound B. The more pessimistic Bound B does not apply to the cases were matrices $\{H^{[i,j]}\}$ are not completely random, such as the cases of diagonal or block-diagonal matrices. Therefore, Bound B does not apply to the cases of time-extension and MIMO-OFDM.

*B. Proof in [6], Dimensions and Mappings*

Bresler, Cartwright and Tse gave another proof for the cases when the $\{H^{[i,j]}\}$ matrices are non-sparse [6]. This section presents our interpretation of the proof presented in [6] in more detail. This understanding forms the basis of the discussions and proposals in subsequent sections.

1) *Proof Outline*

The basic idea is considering equation (5) in two ways. One may fix the $\{U^{[i]}, V^{[i]}\}$ and solve for $\{H^{[i,j]}\}$, or vice versa.

When solving for $\{H^{[i,j]}\}$, equation (5) is a system of $N_e$ *linear* equations. Therefore, given a particular collection of $\{U^{[i]}, V^{[i]}\}$, the dimension of the solution space is $N_H - N_e$. $N_H$ is the number of free elements (i.e., those not confined to zero) in all $\{H^{[i,j]}\}$. A different set of $\{U^{[i]}, V^{[i]}\}$ results in a different solution set. Since the entire choice of $\{U^{[i]}, V^{[i]}\}$ spans a space of dimension $N_v$, the entire solution space has the dimension of $N_v + N_H - N_e$.

On the other hand, this dimension must not be smaller than $N_H$, which is the dimension of the space spanned by all possible $\{H^{[i,j]}\}$. Otherwise, there will be a subset of $\{H^{[i,j]}\}$ not included in the solution space as described above. That means if a realization of the channels $\{H^{[i,j]}\}$ falls into this subset, there would not be corresponding $\{U^{[i]}, V^{[i]}\}$. Namely, equation (5) would not have solution $\{U^{[i]}, V^{[i]}\}$ given such $\{H^{[i,j]}\}$. This violates the requirement that $\{U^{[i]}, V^{[i]}\}$ is solvable when $\{H^{[i,j]}\}$ is generic.

From the above argument, a necessary condition for the solutions to equation (5) is

$$N_v + N_H - N_e \geq N_H . \qquad (20)$$

This leads to the properness condition (16). A more detailed and rigorous proof along this line is given below.

2) *Starting Theorem*

The following is a recount of the proof of Theorem 1 in [6]. The recount is not mathematically rigorous, for the sake of brevity. However, it highlights the steps that are important to our discussion. For more details in mathematical treatment, refer to [6].

The proof starts with a theorem in algebraic geometry (Theorem 9 in [6]), reproduced as follows:

Theorem IV.1: Let $f: X \mapsto Y$ be a polynomial map between irreducible varieties. Suppose that $f$ is dominant, i.e., its image is dense in $Y$. Let $n$ and $m$ denote the dimensions of $X$ and $Y$ respectively. Then $m \leq n$ and

1. For any $y \in f(X) \subset Y$ and for any component $Z$ of the fiber $f^{-1}(y)$, the dimension of $Z$ is at least $n - m$.
2. There exists a nonempty open subset $U \subset Y$ such that $\dim f^{-1}(y) = n - m \; \forall y \in U$.

This theorem is about two varieties, $X$ and $Y$, with a polynomial map between them. A variety is the space constructed by all zeros of a system of polynomial equations. And a polynomial map is a map expressed in polynomial functions.

---

[1] Note that in equation (5), the equation coefficients are not strictly generic, since the same elements in $\{H^{[i,j]}\}$ appear in multiple polynomial equations. In this case, Bernstein's theorem gives an upper limit to the number of equations. Therefore, the necessity of the proper condition is still valid. For more discussions on Bernstein's theorem and its applicability to the IA problem, see Appendix A of [14].

Note that this theorem links the overall dimension of two varieties or their subsets to the dimension of one particular map. In statement 2), mappings from all points $y \in U$ have the same dimension: $\dim f^{-1}(y)$ is independent of $y$ in the said open subset.

3) *Setting Up the Varieties*

Consider the space for $\{H^{[ij]}\}$, referred to as $\tilde{H}$. In MIMO case (i.e., non-spares channel matrices), this is a vector space of dimension

$$dim\, \tilde{H} = \sum_{i \neq j} N_i M_j. \qquad (21)$$

Each vector in this space contains the values of all elements in $\{H^{[ij]}, i \neq j\}$. Matrices $H^{[ii]}$ do not matter in the present consideration.

Note that $dim\, \tilde{H}$ is the $N_H$ defined in the previous subsection.

Denote the space for $\{U^{[i]}\}$ and $\{V^{[i]}\}$ as $\breve{G}(d_i, M_i)$ and $\breve{G}(d_i, N_i)$, respectively. $\breve{G}$s are Grassmannians, where each element can be represented by a matrix. Define a strategic space $\breve{S}$:

$$\breve{S} = \prod_i \breve{G}(d_i, M_i) \times \prod_i \breve{G}(d_i, N_i) \qquad (22)$$

$\breve{S}$ is the direct product of all $\breve{G}(d_i, M_i)$ and $\breve{G}(d_i, N_i)$. Its points are represented by pairs $\{U^{[i]}, V^{[i]}\}$. The dimension of $\breve{S}$ (as a projective variety) is

$$dim\, \breve{S} = N_v. \qquad (23)$$

Denote a collection of $\{U^{[i]}, V^{[i]}, i = 1,2,\ldots,K\}$ as $s$, which is an element in $\breve{S}$. Similarly, a collection of $\{H^{[ij]}, i, j = 1,2,\ldots,K, i \neq j\}$ is referred to as $h$, which is an element in $\tilde{H}$.

Let $\breve{I}$ be the set of all points $(s, h)$ that satisfy equation (5). It is a subset of $\breve{S} \times \tilde{H}$. $\breve{I}$ is a variety because it contains the solutions for polynomial equations in (5), when both $\{U^{[i]}, V^{[i]}\}$ and $\{H^{[ij]}\}$ are considered as variables. $\breve{S}$ and $\tilde{H}$ are also varieties when $\{U^{[i]}, V^{[i]}\}$ and $\{H^{[ij]}\}$ are considered as variables in (5), respectively.

It can be stipulated (justified in [6]) that these varieties are all irreducible. It is known that if $\breve{X}$ is an irreducible variety and $\breve{U} \subset \breve{X}$ is open, then [15]

$$dim\, \breve{U} = dim\, \breve{X}. \qquad (24)$$

Namely, almost all points in an irreducible variety belong to any open subset.

The key to the proof of properness is considering mapping among the three varieties $\tilde{H}$, $\breve{S}$ and $\breve{I}$. Such mappings are based on equation (5) while different parts of $U, V$ and $H$ are considered as variables.

4) *Map from $\breve{I}$ to $\breve{S}$ and Back*

First, dimension of $\breve{I}$ is established. This is achieved by mapping $\breve{I}$ to $\breve{S}$ (by taking the $s$ part of each element in $\breve{I}$), and considering the fibers (inverse mapping) and applying Theorem IV.1.

In order for the theorem to apply, the mapping must be dominant. This means the image of $I$ is a dense set $\breve{S}'$ in $\breve{S}$. In other words, "almost" all points in $\breve{S}$ is a part of some $(s, h)$ pair in $\breve{I}$. Dominance will be shown to be is true in this case.

The inverse mapping from $\breve{S}$ to $\breve{I}$ is equivalent to the following problem: given a point $s$, find the corresponding $h$ points that are in $I$, i.e., that satisfy equation (5). Namely, in equation (5), $U$s and $V$s are fixed, one solves for the $H$ matrices.

This is a linear algebra problem. Therefore, the dimension of the solutions space is well known. There are a total of $N_e$ equations, and $dim\, \tilde{H}$ variables. Therefore, the dimension of the solution space is

$$dim\, p^{-1}(\breve{I} \mapsto \breve{S}) = dim\, \tilde{H} - N_e. \qquad (25)$$

Because $d_i \leq N_i$ and $d_i \leq M_i$, equations (21) and (14) guarantees, independent of $s$:

$$dim\, \tilde{H} - N_e \geq 0. \qquad (26)$$

This implies mapping $\breve{I} \mapsto \breve{S}$ is dominant. Namely, almost any $s \in \breve{S}$ can yield some solution $h$ from (5), and therefore belongs to some pairs of $(s, h) \in \breve{I}$. This is not true, as will be pointed out in the next subsection, in certain cases of sparse $\{H^{[i,j]}\}$ matrices.

Using statement 2) of Theorem IV.1 while equating an open subset in a variety and the variety itself (in the sense of probability-1), the following equation is established:

$$dim\, p^{-1}(\breve{I} \mapsto \breve{S}) = dim\, \breve{I} - dim\, \breve{S}. \qquad (27)$$

Combining equations (23), (25) and (27),

$$dim\, \breve{I} = dim\, p^{-1}(\breve{I} \mapsto \breve{S}) + dim\, \breve{S}$$
$$= dim\, \tilde{H} - N_e + dim\, \breve{S} = dim\, \tilde{H} - N_e + N_v. \qquad (28)$$

5) *Map from $\breve{I}$ to $\tilde{H}$ and Back*

Now focus on the inverse mapping dimension from $\tilde{H}$ to $\breve{I}$, which is the dimension of the solutions space of equation (5), given $h \in \tilde{H}$.

In order to apply theorem IV.1, the mapping needs to be dominant. This is ensured by *assuming* (justified because necessary condition is the issue under consideration) generic system has solutions. Namely, "almost" any point in $\breve{H}$ would yield at least one solution $s \in \breve{S}$. Therefore, almost any point in $\breve{H}$ belongs to the image of $\breve{I}$.

Using statement 2) of Theorem IV.1 one more time:

$$dim\, p^{-1}(\breve{I} \mapsto \breve{H}) = dim\, \breve{I} - dim\, \breve{H}\,. \qquad (29)$$

Combining equation (29) with equation (28):

$$dim\, p(\breve{H} \mapsto \breve{I}) = dim\, p^{-1}(\breve{I} \mapsto \breve{H}) = N_v - N_e\,. \qquad (30)$$

The necessary condition for the solution to equation (5) to exist is $dim\, p(\breve{H} \to \breve{I}) \geq 0$. Therefore, properness requirement (16) is derived. Remember this is for MIMO case, where inequality (26) holds.

6) *The Case of Diagonal Channel Matrices*

Now consider the case where all matrices in $\{H^{[i,j]}\}$ are required to be diagonal. Each matrix thus has only $N_i$ non-zero variables. Equation (21) becomes

$$dim\, \breve{H}_d = \sum_i N_i\,. \qquad (31)$$

Here $\breve{H}_d$ is the space spanned by all none-zero variables in $\{H^{[i,j]}\}$. Therefore, in many cases

$$dim\, \breve{H}_d \leq N_e = \sum_{k \neq l} d_k d_l\,. \qquad (32)$$

Inequality (26) is no longer true. Therefore, the proof in [6], as outlined in the previous subsections, does not apply.

In this situation, not all $s \in \breve{S}$ points can be mapped to an $h \in \breve{H}_d$, when $h$ is considered as variable to be solved. Only those $s$ points that make some of the equations in (5) become linearly dependent (thus reducing the total number of independent equations) can yield non-trivial solutions in $\breve{H}_d$. Therefore, mapping $\breve{I} \to \breve{S}$ is not dominant. Theorem IV.1 is thus no longer applicable. The same argument applies to the case of MIMO-OFDM where $\{H^{[i,j]}\}$ are required to be block-diagonal.

## V. POSSIBLE WAYS FORWARD

As pointed out in Sections IV, so far the proofs of the properness condition (17), outlined in Sections IV.A and IV.B, do not apply when the channel matrix is sparse (such as diagonal or block-diagonal channel matrices). On the other hand, Bound A (Section III.A) applies only when signal dimension tends to infinity. Therefore, the actual DoF bounds in the cases of sparse channel matrices (i.e., with time extension and MIMO-OFDM) are still unknown. The goal, therefore, is to derive a better bound in the case of sparse channel matrices (particularly diagonal and block-diagonal matrices) with limited dimension.

The same goal has been recognized by [14]. That work started from the case of single beams (i.e., each user sends only one data stream) and studied the bound for total number of users (equivalent to the total number of DoF). Several interesting and strong results were obtained. The framework and results are intended to be extended to the more general multi-beam cases.

In this section we discuss a few possible approaches, based on the approach of [6].

### A. Theoretical Approach

As pointed out in Section IV.B, the dimension analysis performed in [6] does not apply in the cases of spares channel matrices, because the dimension of $\breve{H}_d$ space is too small, see inequality (32). Here $\breve{H}_d$ is the space spanned by all non-zero variables in $\{H^{[i,j]}\}$. This section extends such discussion, following the notations therein.

Let $\breve{S}_d$ be the set of such $s$ that yields solutions of equation (5) for any $h \in \breve{H}_d$. $\breve{S}_d$ is a subset of $\breve{S}$, which is the set of all possible values in $\{U^{[i]}, V^{[i]}, i = 1, 2, \ldots, K\}$.

$$dim\, \breve{S}_d \leq dim\, \breve{S} = N_v\,. \qquad (33)$$

To find some information about $dim\, \breve{S}_d$, Equation (5) can be written as

$$Ph = 0\,. \qquad (34)$$

Here $h$ is a vector of size $dim\, \breve{H}_d$, collecting all non-zero components of $\{H^{[i,j]}\}$; $P$ is a matrix of size $N_e \times dim\, \breve{H}_d$. $P$ is a function of $\{U^{[i]}, V^{[i]}\}$. Therefore, $dim\, \breve{S}_d$ equals to the number of free variables in $P$.

Obviously, many $P$ matrices generate the same $h$. The distinctions among them are not important to the current problem. Therefore, they can be grouped together. Given $h$, define set $\breve{P}_h$ so that

$$Ph = 0 \,\, \forall P \in \breve{P}_h\,. \qquad (35)$$

For a given $h$, all row in $P \in \breve{P}_h$ must lay in a subspace of dimension $dim\, \breve{H}_d - 1$ that is orthogonal to $h$. Such subspace has one-to-one correspondence with $h$ (up to a scalar factor, which can be ignored). Therefore, set $\breve{P}_h$ can be represented by $(dim\, \breve{H}_d - 1)$ linearly independent bases of the subspace. Namely, one can represent $\breve{P}_h$ with a full-rank matrix $\bar{P}_h$ of size $dim\, \breve{H}_d - 1$ by $dim\, \breve{H}_d$. Note that another matrix $\bar{\bar{P}}_h$ represents the same $\breve{P}_h$, if

$$\bar{\bar{P}}_h = Q\bar{P}_h\,. \qquad (36)$$

Here $Q$ is a full rank matrix of size $\dim \breve{H}_d - 1$ by $\dim \breve{H}_d - 1$. Matrix $Q$ can be chosen so that

$$\bar{\bar{P}}_h = \begin{pmatrix} I_{\dim \breve{H}_d - 1} & \bar{\bar{p}}_h \end{pmatrix}. \tag{37}$$

Here $I_{\dim \breve{H}_d - 1}$ is an identity matrix, and $\bar{\bar{p}}_h$ is a vector of size $\dim \breve{H}_d - 1$. Therefore, $\bar{\bar{p}}$ has a one-to-one mapping to $\check{P}_h$. Namely, the maximum dimension of $\{\bar{\bar{P}}\}$, which is $\dim \check{S}_d$, is $\dim \breve{H}_d - 1$. On the other hand, not all vectors $\bar{\bar{p}}$ are permissible. This is because $P$ is constructed from the subset $\check{S}_d$. Its elements may be constrained. Therefore, its dimension may be lower:

$$\dim \check{S}_d \le \dim \breve{H}_d - 1 \,. \tag{38}$$

Comparing to Section IV.B, $\dim \breve{H}_d - 1$ plays the role of $N_e$, and $\dim \check{S}_d$ plays the role of $N_v$. Therefore, based on (38), (30) becomes

$$\dim p \,(\breve{H}_d \mapsto I) = \dim \check{S}_d - \dim \breve{H}_d - 1 \le 0 \,. \tag{39}$$

(39) shows that the dimension of solution is at most 0. This means that if solutions exist, it is in a zero-dimensional space. Namely, the number of solution is finite, disregarding the extra freedoms associated with (36). Furthermore, this happens when the equality in (38) holds.

Therefore, whether equation (5) has solution depends on the dimension of $\check{S}_d$. Unfortunately it is not easy to determine unless one can construct $\check{S}_d$ short of solving equation (5) directly. On the other hand, the fact that (5) has at most finite number of solutions, as implied by (40), may be helpful in selecting some algebraic geometric tools for further analysis.

### B. Numerical Probing

Another approach is using numerical methods to probe the solution space of (34). More specifically, vectors $h$ can be solved from the linear equation (34), given $\{U^{[i]}, V^{[i]}\}$. Such solutions can be accumulated by choosing $\{U^{[i]}, V^{[i]}\}$ randomly (some of them do not yield non-trivial solutions). If the accumulated solutions fill the whole space $\breve{H}_d$, it can be expected that any $h \in \breve{H}_d$ would have at least one corresponding $\{U^{[i]}, V^{[i]}\}$. Namely, equation (5) has solution for any given $h$.

However, the word "fill" above is not well defined. Ideally, "fill" would mean that the accumulated $\{h\}$ forms a subset of $\breve{H}_d$, and is measure 1. Unfortunately, such "fill" is not easy to test based on a finite number of trials (with random $\{U^{[i]}, V^{[i]}\}$).

One possible way is examining the dimension of the space spanned by the solutions $\{h\}$. If it is the same as $\breve{H}_d$, it might be an indication that the solutions "fill" the whole space. Note that this has not been mathematically proven. Namely, if $h_1 \in \breve{H}_d, h_2 \in \breve{H}_d$ both yield solutions for equation (5), a linear combination $\alpha h_1 + \beta h_2$ does not necessarily yield solution. Nevertheless, the dimension may be a good indicator. Its efficacy can be assessed by comparing the results of some special cases, where the existence and nonexistence of solutions are known.

### C. Numerical Solutions

Another approach of assessing DoF upper bound is directly solving (5) for generic channel matrices (with the proper constrains of sparsity). Multiple trials would show whether non-trivial solutions can be obtained with a probability of 1.

Equation (5) is a form of bilinear equation. Solution to bilinear equations has been researched for many years and so far there is no effective algorithm. Although (5) is a special form where a majority of the terms are missing in the equations, there is no known solution algorithm in the literature.

In general, solving polynomial equations numerically is not easy. One promising method may be the Gröbner basis theory. This theory provides an algorithm that reduces the number of polynomials in the system and eventually determines whether a nontrivial solution exists before solving for it. Although Gröbner basis algorithm is guaranteed to finish in finite steps, in general there is no upper bound on time and memory needed to solve a particular problem. It was hoped that since IA problem (5) is only second order and the number of monomials in each equation is relatively small, the algorithm would work efficiently. However, this needs to be verified by experiments. There are public domain Gröbner basis software package available. Therefore, it is advisable to conduct some test to assess the feasibility of such approach.

In addition to attempting to derive exact solutions, there are numerical algorithms to produce approximate solutions [9-13]. This may be another way to accumulate observations that will guide future work. The algorithms demand much less resources (CPU time and memory) than the Gröbner basis technique. However, it is not clear what level of accuracy (i.e., residual mutual interference) is required in order to be connected with the analytical considerations.

## VI. SUMMARY

This paper provided an overview of the current state of the arts in determining DoF upper bounds for IA systems. It has been shown that the current proofs of the properness condition (17) do not appear to apply to the cases where the channels are diagonal or block-diagonal, which is practical in OFDM and MIMO-OFDM IA systems. On the other hand, for diagonal channels, a tight upper bound is known only when the IA dimension approaches infinity.

Therefore, it is premature to believe that the properness bound (17) limits performance of IA systems based on OFDM and MIMO-OFDM technologies. Establishing a suitable upper bound for such systems is of significant theoretical and practical importance. A few approaches have been suggested in Section V.

ACKNOWLEDGMENT

The author wishes to thank H. Chen of University of California, Berkeley for very valuable discussions. Such discussions help to build the understandings reflected in Section IV.

The author also wishes to thank R. Sun of University of Minnesota, for discussions concerning his work in [14].

Discussions and collaborations from Dr. S. Chin and L. Liu of Johns Hopkins University are gratefully acknowledged.